\documentclass[twoside,10pt]{article}
\usepackage{epsfig}

\makeatletter
\def\modxspace{\futurelet\@let@token\@modxspace}
\let\@modx@space\space
\def\@modxspace{%
  \ifx\@let@token\bgroup\else
  \ifx\@let@token\egroup\else
  \ifx\@let@token\/\else
  \ifx\@let@token\ \else
  \ifx\@let@token~\else
  \ifx\@let@token.\else
  \ifx\@let@token!\else
  \ifx\@let@token,\else
  \ifx\@let@token:\else
  \ifx\@let@token;\else
  \ifx\@let@token?\else
  \ifx\@let@token'\else
  \ifx\@let@token)\else
  \ifx\@let@token-\else
   \@modx@space
   \fi\fi\fi\fi\fi\fi\fi\fi\fi\fi\fi\fi\fi\fi}
\def\noxspace#1{{\let\@modx@space\null#1}}
\makeatother

\def\meanPt{$\left<p_T\right>\:$}
\newcommand{\be}{\begin{equation}}
\newcommand{\ee}{\end{equation}}
\newcommand{\bea}{\begin{eqnarray}}
\newcommand{\eea}{\end{eqnarray}}

\newcommand{\dedx}{\ensuremath{\frac{dE}{dx}}\modxspace}

\newcommand{\lam}{\ensuremath{\Lambda}\modxspace}

\newcommand{\alam}{\ensuremath{\overline{\Lambda}}\modxspace}

\newcommand{\meanpt}{\mbox{\ensuremath{<\!\!p_{t}\!\!>}}}

\newcommand{\pbar}{\ensuremath{\overline{p}}\modxspace}

\newcommand{\kz}{\ensuremath{K^{0}_{S}}\modxspace}

\newcommand{\pp}{\ensuremath{\pi^+}\modxspace}
\newcommand{\pmin}{\ensuremath{\pi^-}\modxspace}

\newcommand{\rap}{\ensuremath{{\bf y}}\modxspace}
\newcommand{\pt}{\ensuremath{p_{T}}\modxspace}
\newcommand{\mt}{\ensuremath{m_{T}}\modxspace}

\newcommand{\dndy}{\ensuremath{dN/d{\bf y}}\modxspace}

\begin{document}
\topmargin-2.8cm
\oddsidemargin-1cm
\evensidemargin-1cm
\textwidth18.5cm
\textheight25.0cm

\title{\vspace{1cm} Neutral strange particle production at mid unit rapidity \\in $p+p$ collisions at $\sqrt{s}$ = 200 GeV}
\author{J.\ Adams$^1$ and M.\ Heinz$^2$  for the STAR collaboration
\\
\\
$^1$University of Birmingham, UK
\\
$^2$University of Bern, Switzerland}

\maketitle

\begin{abstract} 
    
We briefly discuss the methods of analysing reconstructed neutral strange particles in $p+p$ collision data measured at $\sqrt{s}$ = 200 GeV taken using the Solenoidal Tracker At RHIC (STAR) detector.  We present spectra for \kz, \lam and \alam as a function of \pt and multiplicity, and compare to previous high energy $p+\pbar$ collision data.  The two component nature of the spectra suggests contributions from hard and soft processes, and the observed increase of \meanpt $\:$ with multiplicity indicates a growing contribution from the hard processes.  The work described herein was presented as a poster at the Quark Matter 2004 conference.   
\end{abstract}

\section{Introduction}

Particles which contain strange quarks are valuable probes of the dynamics of $p+p$ collisions, as constituent strange quarks are not present in the initial colliding nuclei.  The relative enhancement of the strange particle yield per participant nucleon from $p+p$ collisions to heavy ion collisions has been suggested as a possible Quark Gluon Plasma signature \cite{Rafelski}. The strangeness yield in $p+p$ collisions is particularly important because it forms the baseline from which any possible strangeness enhancement in Au+Au collisions is determined.  We report the first measurements of inclusive neutral strange particle production in $p+p$ collisions at a centre of mass energy ($\sqrt{s}$) of 200 GeV.  The data was acquired during the 2001-2002 run and approximately 10 million events were available for analysis.  We compare the \lam, \alam and \kz spectra with previous $p+\pbar$ collision data taken at $\sqrt{s}$ between 200 and 1800 GeV.    
\\

\section{Analysis}

The STAR experiment consists of a number of detectors and is described in detail elsewhere \cite{STAR1},\cite{STAR2}.   Neutral strange particle decays, v0s, were identified by analysing the tracks made by their charged daughter particles.  The following decay channels were used in order to reconstruct \lam, \alam and \kz candidates:
\begin{center}
 $\lam \rightarrow p + \pmin \: (69\%) \quad \quad \: \alam \rightarrow \pbar + \pp \:  (69\%)  \quad\quad\: \kz \rightarrow \pp + \pmin \: (64\%)$
\end{center}
\noindent Daughter particle identification was achieved by analysing the specific ionisation energy loss (\dedx) of tracks.  Combinatoric background mainly formed by the random crossings of oppositely charged tracks was reduced by the application of topological cuts (e.g a distance of closest approach of the v0 candidates daughter tracks).  In the analysis corrections were made for the primary vertex finding and v0 finding efficiencies by embedding Monte Carlo tracks into real data.  Using a previous measurement of the $\Xi^-$ yield it was possible to make an estimate of the feed down correction for \lam and \alam.
\\
\newpage
\section{Study of yield and \meanPt}

\begin{figure}
\begin{center}
\epsfig{figure=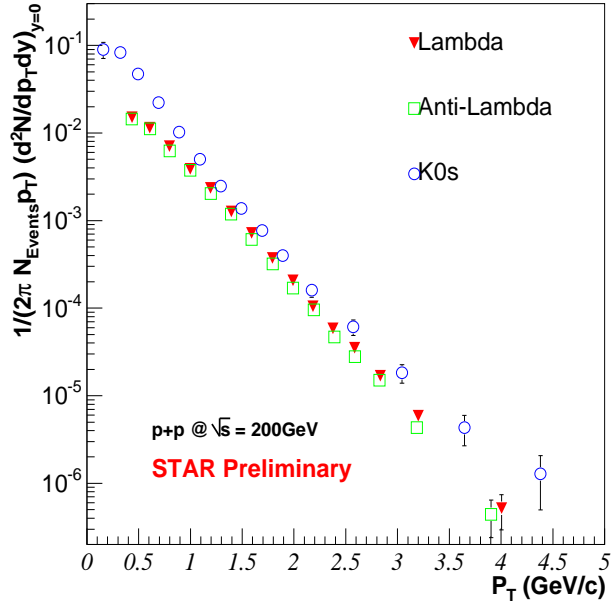, height=8cm, width=8cm}
\caption{Minimum-bias, non-feed down corrected,  \lam, \alam, and \kz spectra with ($\mid{y}\mid \leq 0.5$) from $p+p$ at $\sqrt{s} = 200 $GeV.  The errors in the plot are statistical only.}
\label{fig:AllSpec}
\end{center}
\end{figure}

Non-feed down corrected, minimum bias trigger, spectra for \lam, \alam and \kz, are superposed in Figure \ref{fig:AllSpec}.  As the fiducial region of the TPC limits the \pt acceptance at mid-rapidity to greater than 0.3 GeV (\lam) and greater than 0.1 GeV (\kz), it is necessary to fit the data and extrapolate the fit function in order to determine total particle yields and \meanPt.
\\

\noindent Previous $p+\pbar$ experiments UA5\cite{UA5} and UA1\cite{UA1} used exponentials in $m_T$ and $p_T$, and power law functions to fit the strange spectra.  With this higher statistics measurement we have found that none of these functions alone is able to describe the data well over the whole \pt range.  A better parameterisation is obtained with a two component fit.  An exponential function in \mt provides a good description of the data at \pt $\leq$ 1 GeV, where as at high \pt  an exponential in $p_{T}$ (\lam,\alam) or a power law (\kz) describes the data.  Therefore composite fits, represented in equations \ref{Eq:ExpmtandExppt} (for \lam, \alam) and \ref{Eq:ExpandPower} (for \kz), were applied to the data shown in Figure \ref{fig:AllSpec}.  The \meanPt $\:$ and \dndy were extracted by integrating the fits over all \pt and are tabulated in table \ref{tab:Results1}.  A full study of the systematic errors is in progress but we estimate an uncertainty in the range of $10\%$.  The errors quoted in table \ref{tab:Results1} are the variations in \dndy and \meanPt $\:$ when individual exponentials in $m_T$ and $p_T$ and power laws are fitted to the spectra. 
\\

\begin{equation}
\frac{1}{2\pi p_{T}}\frac{d^{2}N}{d\rap dp_{T}} = A.e^{\frac{-m_{T}}{T}} + B.e^{\frac{-p_{T}}{T}}
\label{Eq:ExpmtandExppt}
\end{equation}

\begin{equation}
\frac{1}{2\pi p_{T}}\frac{d^{2}N}{d\rap dp_{T}} = C.e^{\frac{-m_{T}}{T}} + D.(1+\frac{p_{T}}{p_{0}})^{-n}
\label{Eq:ExpandPower}
\end{equation}

\begin{table}[ht]
\begin{center}
\begin{tabular}{|c|c|c|c|c|c|c|c|} \hline
Particle&\dndy&\dndy in measured& \dndy feed & \meanPt \\ 
&&\pt region $\pm$ stat&down corrected&\\ \hline
\lam & 0.044 $\pm$ 0.003 &0.038 $\pm$ 0.001 &0.034 $\pm$ 0.005 &  0.76 $\pm$ 0.05 \\ \hline
\alam & 0.042 $\pm$ 0.003  & 0.036 $\pm$ 0.001&0.032 $\pm$ 0.005 & 0.75 $\pm$ 0.05 \\ \hline
\kz &0.123 $\pm$ 0.006  & 0.115 $\pm$ 0.001&& 0.60 $\pm$ 0.05 \\ \hline
\end{tabular}
\caption[\dndy and \meanPt]{A summary of \dndy and \meanPt $\:$for neutral strange particles in $p+p$ at $\sqrt{s}$ = 200 GeV with $\mid$y$\mid$ $\leq$ 0.5.  The errors quoted are statistical + systematic, where the systematic error is estimated from the results of different fits to the data}
\label{tab:Results1}
\end{center}
\end{table} 

\begin{table}[ht]
\begin{center}
\begin{tabular}{|c|c|c|c|c|c|} \hline
Particle&STAR \dndy&UA5 \dndy&UA5 \dndy  \\ 
&$\mid$y$\mid$ $\leq$ 0.5&$\mid$y$\mid$ $\leq$ 2.0&$\mid$y$\mid$ $\leq$ 0.5\\ \hline
\lam + \alam & 0.066 $\pm$ 0.004 & 0.27 $\pm$ 0.07 &  0.08 $\pm$ 0.02 \\ \hline
$\frac{\lam+\alam}{2\kz}$ & 0.27 $\pm$ 0.04 & 0.31 $\pm$ 0.09 & 0.31 $\pm$ 0.09 \\ \hline
\end{tabular}
\caption[Comparison of \dndy] {A summary of \dndy for \lam + \alam and $\frac{\lam+\alam}{2\kz}$ (\lam, \alam feed down corrected) measured by the UA5 \cite{UA5} and STAR experiments at $\sqrt{s}$ = 200 GeV}
\label{tab:Results2}
\end{center}
\end{table}
\begin{table}[ht]
\begin{center}
\begin{tabular}{|c|c|c|c|} \hline
Particle& STAR \meanPt&UA5 \meanPt \\ 
&$\mid$y$\mid$ $\leq$ 0.5&$\mid$y$\mid$ $\leq$ 2.0\\ \hline
\lam + \alam & 0.76 $\pm$ 0.05 & 0.8 +0.2,-0.14\\ \hline
\kz & 0.60 $\pm$ 0.05 & 0.53 +0.08,-0.06\\ \hline
\end{tabular}
\caption[Comparison of \meanPt] {A summary of \meanPt $\:$for \lam + \alam (feed down corrected) and \kz  measured by the UA5 \cite{UA5} and STAR experiments at $\sqrt{s}$ = 200 GeV}
\label{tab:Results3}
\end{center}
\end{table}

\noindent The UA5 experiment \cite{UA5} measured \lam, \alam and \kz in $p+\pbar$ collisions at $\sqrt{s}$ = 200 GeV.  In order to compare \dndy over the STAR rapidity range $(\mid{y}\mid\leq 0.5)$, the UA5 $(\mid{y}\mid\leq 2.0)$ yield has been scaled according to the STAR rapidity interval, using the rapidity distribution for \lam, \alam and \kz generated from a Pythia simulation \cite{Pythia}.  A comparison of the \dndy is tabulated in table \ref{tab:Results2} and a comparison of \meanPt $\:$in table \ref{tab:Results3}.  The \dndy and \meanPt $\:$measured by STAR and UA5 agree within the quoted errors.  This indicates that at mid-rapidity and $\sqrt{s}$ = 200 GeV, particle production does not depend greatly on the net baryon number of the colliding nuclei.
\\

\section{\meanPt $\:$ vs Multiplicity}

\begin{figure}
\begin{center}
\epsfig{figure=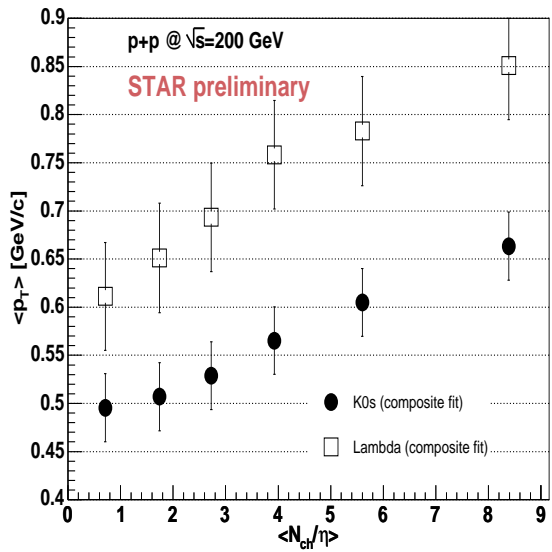, height=8cm, width=8cm}
\caption{\meanPt $\:$vs charged uncorrected multiplicity per unit pseudo rapidity, (uncorr $Nch/\eta$) obtained with composite function fits. A hardening of the spectra is observed with increasing event multiplicity.  The errors in the plot are statistical only.}
\label{fig:ptvM}
\end{center}
\end{figure}

\noindent Figure \ref{fig:ptvM} shows that the \meanPt $\:$of the \lam and \kz increases with measured (uncorrected) track multiplicity.  This result is similar to the trend observed by UA1 \cite{UA1} and E735 \cite{E735} and is consistent with the idea of the \meanPt $\:$  and multiplicity increasing with the momentum transfer of the parton-parton collisions.  The correlation between \meanPt$\:$ and multiplicity may be due to increasing strange particle production from harder processes such as jet and mini-jet fragmentation mechanisms \cite{WangGyulassy}. 
\\

\begin{figure}
\begin{center}
\epsfig{figure=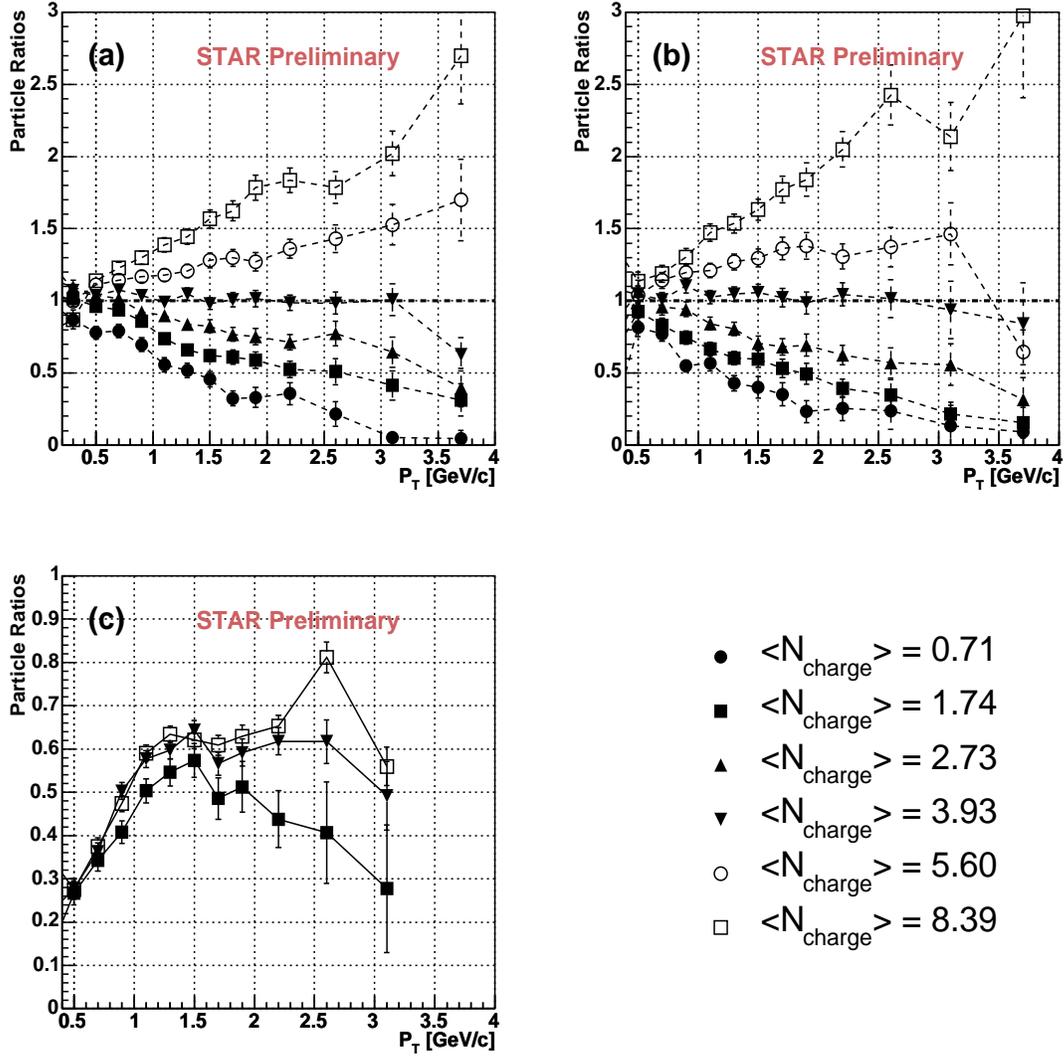, height=15cm, width=15cm}
\caption{Spectra for \kz (a) and  \lam (b) normalised by the min-bias distribution for different uncorrected multiplicity classes.  Panel c shows the ratio of \lam / \kz in the lowest (filled squares) and highest multiplicity bin (open squares) as well as the ratio of the min bias result (filled triangles).  The errors are statistical}
\label{fig:MultClass}
\end{center}
\end{figure}

\noindent In Figure \ref{fig:MultClass} we confine the spectra into six multiplicity classes and divide by the minimum-bias spectrum.  From Figures \ref{fig:AllSpec} and \ref{fig:MultClass} we can infer that for all event multiplicity classes the low \pt part of the spectra dominates the yield, indicating that soft particle production occurs readily for all types of events which can be characterised by event multiplicity.  The correlation between \meanPt $\:$and multiplicity shown in Figure \ref{fig:ptvM} is driven by the low \pt ($\pt < 1.5$ GeV) particles as panels a) and b) in Figure \ref{fig:MultClass} show an increase in the low \pt region as multiplicity increases.  For low multiplicity classes there is a large deficit of high \pt particles and for high multiplicity classes, where there is more momentum transfer, there is a large enhancement.
\\

\noindent Panel c in Figure \ref{fig:MultClass} shows the \lam to \kz ratio for high, minimum-bias and low multiplicity classes.  It indicates that at high \pt, \lam production is relatively more favoured over \kz for high multiplicity events, and that \kz production is relatively favourable for low multiplicity events.  This is perhaps because the \lam, being more massive than the \kz, is harder to produce in low multiplicity collisions, where there is less momentum transfer.

\section{Conclusions and future work}

The STAR experiment has made the first high statistics measurements of mid-rapidity \lam, \alam, and \kz generated in $p+p$ collisions at $\sqrt{s}$ = $200$ GeV.  The measured \meanPt $\:$and \dndy agree with those measured by the UA5 collaboration for $p+\pbar$ at $\sqrt{s}$ = $200$ GeV.  We have shown that the \alam yield is very similar to the \lam yield which indicates that there is small net baryon number at mid-rapidity.  We observe an increase in the \meanPt $\:$with measured multiplicity for \lam and \kz, which indicates a growing contribution from harder processes.  Furthermore the multiplicity integrated spectra are best described by a two component fit. This implies that the particle production mechanism is one which produces significant quantities of soft strange hadrons, best described by an exponential in \mt as well as higher \pt strange hadrons, which are best described by power law or \pt exponential functions.  A study of the systematic errors and a more thorough feed down correction for \lam and \alam is underway.
\\

\begin{thebibliography}{99}

\bibitem{Rafelski}J.Rafelski and B. Muller, Phys. Rev. Lett. 48, 1066 (1982)
\bibitem{STAR1}K.H.Ackermann et al, (STAR Collaboration), Nucl. Phys. A661, 681c (1999)
\bibitem{STAR2}K.H.Ackermann et al, (STAR Collaboration), Nucl. Instrum. Meth. A499, 624 (2003)
\bibitem{UA5}R.E.Ansorge et al, (UA5 Collaboration) Nucl. Phys. B328, 36 (1989)
\bibitem{UA1}G. Bocquet et al (UA1 Collaboration), Phys. Lett. B 366, 441 (1996)
\bibitem{Pythia}T. Sjostrand et al. Pythia 6.2, Physics and Manual, hep-ph/0308153 (2002)
\bibitem{E735}T.Alexopolous et al (E735), Phys. Rev. D Vol 48 , 3, 984 (1993)
\bibitem{WangGyulassy}X. Wang, M. Gyulassy, Phys. Lett. B 282, 466 (1992)

\end{thebibliography}
\end{document}